\documentclass[12pt]{article}
\newcommand{\beq}{\begin{equation}}
\newcommand{\eeq}{\end{equation}}
\newcommand{\bea}{\begin{eqnarray}}
\newcommand{\eea}{\end{eqnarray}}

\newcommand{\half}{\frac{1}{2}}

\usepackage{graphicx}
 \usepackage{setspace}
 \usepackage{amsfonts}
 \usepackage{amsmath}  
 \usepackage{txfonts} 
\begin{document}
\pagestyle{empty}
\begin{center}
\vskip 2cm
{\bf \LARGE (1+1)-Dimensional Entropic Gravity}
\vskip 1cm
{\large R. B. Mann\footnote{Email: rbmann@sciborg.uwaterloo.ca}\\
{\it Department of Physics and Astronomy, University of Waterloo\\ \& Perimeter Institute for Theoretical Physics}\\
J. R. Mureika\footnote{Email: jmureika@lmu.edu} \\
{\it Department of Physics, Loyola Marymount University}}
\end{center}

{\noindent{\bf Abstract} \\
We consider the formulation of entropic gravity  in two spacetime dimensions. The usual gravitational force law is derived even in the absence of area, as normally required by the holographic principle.  A special feature of this perspective concerns the nature of temperature and entropy defined at a point. We argue that the constancy of the gravitational force in one spatial dimension implies the information contained at each point in space is an internal degree of freedom on the manifold, and furthermore is a universal constant, contrary to  previous assertions that entropic gravity in one spatial dimension is ill-defined.  We give some heuristic arguments for gravitation and information transfer constraints within this framework, thus adding weight to the contention that spacetime and gravitation might be emergent phenomena.
}

\vskip 1cm
\section{Introduction}

The duality between gravitation and thermodynamics is well-known, stemming from the original considerations of Beckenstein and Hawking to black hole evaporation \cite{beckenstein,hawking}.  These pioneering  investigations demonstrated that the entropy of a black hole is described completely in terms of its horizon area, and hence its temperature as a function of the Schwarzschild radius.   The frame-dependence of temperature was noted in the celebrated Unruh effect \cite{unruh}, which implies  a uniformly accelerated detector in a vacuum experiences a thermal bath of radiation.   The gravity-thermodynamics connection was further elucidated in \cite{jacobson}, where it was pointed out that the Einstein field equations may be understood as a collective equation of state.   The most striking instantiation of  gravitational duality is the AdS/CFT correspondence conjecture \cite{maldacena}, which associates a $d-$dimensional conformal field theory with a gravitational theory in one higher dimension.  The generalization of this duality is embodied by the holographic principle \cite{susskind,bousso}, which posits that the entropy content of any region of space is defined by the bounding area of the region.  

Recently, a new perspective on gravitational holography  proposing that the laws of gravitation are no longer fundamental, but rather emerge naturally from the second law of thermodynamics as an ``entropic force'' \cite{padman,verlinde}, has received much attention.   This framework has since been extended to numerous situations, ranging from quantum gravity \cite{smolin,modesto,zhao} and quantum information \cite{myung,lee2,lee3} to cosmological implications \cite{cosmo1}, including implications for black hole temperature \cite{bh1,bh2,bh3,ohta2}, dark energy \cite{smoot1,smoot3,lee,danielsson,Li:2010cj,ohta1}, and  inflation \cite{smoot2,cai,Li:2010bc,Wang:2010jm}.   Beyond Newtonian and relativistic gravity, the entropic formalism has been studied in the context of MOND \cite{mond1,mond2}, $f(R)$ theories \cite{fr}, and even potential connections to Lifshitz gravity \cite{horava}, non-commutative geometry and unparticle physics \cite{piero}.  An interesting consequence of applications arises in the quantum regime, where uncertainty principle constraints applied to information transfer between the test mass and holographic screens necessitate a lower-bound to the mass of the photon and/or graviton \cite{jrmrbm}.  While the predicted mass is within experimental constraints, this provides a robust test of the proposed mechanism.  Additionally, a potential entropic formulation for circular motion has recently been proposed \cite{doug1}.

Being fundamentally based on the holographic principle, one expects that entropic gravity can be generalized to any spatial dimension \cite{verlinde}.  A problematic exception potentially arises for $d=1$, in which there is no concept of area and hence no intuitive extension (or rather subtension) of the area-entropy law.  We consider here the problem of formulating  entropic gravity in one spatial dimension and demonstrate that its resultant physics is well-defined, despite the non-existence of the prerequisite bounding area.   Due also in part to the resurgence in importance of $(1+1)-$dimensional physics in high energy physics ({\it e.g} in models such as causal dynamical triangulations \cite{cdt}, non-commutative geometry \cite{piero1}, or evolving dimensions \cite{dejan1,jrmds}), a complete description from the entropic view may provide additional insights on emergent phenonema.

Section~\ref{section1} provides a review of the entropic gravity framework for a general number of dimensions $d$.  In Section~\ref{section2} we consider the $d=1$ case, in which the information (and entropy) become intrinsic degrees of freedom of the space itself, and comment on the implications of this new interpretation.   We comment on the feasibility of this framework, conjecture potential relativistic extensions of the model, and outline conclusions and future directions in Section~\ref{concspec}.

\section{Entropic Gravity}
\label{section1}
An entropic force may be defined in a purely classical and general sense as \cite{padman,verlinde}
\beq
F_{\rm entropic} \equiv -\frac{\Delta E}{\Delta x} = -T \frac{\Delta S}{\Delta x}~~.
\label{entforce}
\eeq
By definition, $F_{\rm entropic}$ is a force resulting from the tendency of a  system to increase its entropy.   Since $\Delta S >0$, the sign of the force --whether it is repulsive or attractive -- is determined by how one chooses the definition of $\Delta x$ as it relates to the system in question\footnote{In the framework presented in \cite{verlinde}, this force is necessarily attractive, since by design the value of $\Delta x$ is negative (and of course $\Delta S >0$).}.

A mass distribution $M$ induces a holographic screen $\Sigma$ at some distance $r$ that has encoded on it gravitational information.  Consider the situation in $d$ spatial dimensions. According to the holographic principle, the screen encodes all physical information contained within its volume in  bits on the screen whose number $N$ is given by
\beq
N = \frac{A_\Sigma(r)}{\ell_P^{d-1}} 
\label{Narea}
\eeq
where 
\beq
A_\Sigma(r)  =  \frac{2\pi^{d/2}}{\Gamma(\frac{d}{2})}  \; r^{d-1}
\eeq
is the area of the hyperspherical screen and $\ell_P $ is a fixed length scale whose area
 $\ell_P^{d-1}$ is the minimal area containing a single bit.
Assuming that the total energy of the system $E=Mc^2$ is evenly distributed over the bits\footnote{We note assuming {\it a priori} such an equipartition is not necessary.  It has been shown {\it e.g.} that an analogous statistical interpretation of gravitation can be derived purely from black hole spacetimes, with equations of motion resulting from extremization of the entropy \cite{Majhi1,Majhi2}.}
\beq
E = \frac{N}{2} k_b T
\label{equip}
\eeq
one can then eliminate the inferred temperature $T$ in terms of the mass $M$ and area $A_\Sigma(r)$, yielding
\beq \label{kbT}
k_B T = \frac{2Mc^2 \ell^{d-1}_P}{A_\Sigma(r)}
\eeq

A second test mass $m$ will begin to ``transfer'' its own information bits to the screen, a measure of which is taken to be
\beq
\Delta S = 2\pi k_B \frac{\Delta x}{\lambdabar}~~~,~~~\lambdabar = \frac{h}{mc}
\label{entrans}
\eeq
as the particle moves a distance $\Delta x$ toward  the screen.  Note that this yields an attractive force from Equation~\ref{entforce}, since by construction $\Delta x < 0$.   When the particle is within a distance equal to its own Compton length, $\Delta x = \lambdabar$, the particle itself is ``indistinguishable'' from the screen and their bits merge. 

Combining Equations (\ref{kbT},~\ref{entrans}) with Equation (~\ref{entforce}),  it is straightforward to show that the entropic force yields Newton's law of gravitation 
\beq
F_{\rm entropic}= -2\pi^{1-\frac{d}{2}} \Gamma\left(\frac{d}{2}\right)\frac{c^3 \ell^{d-1}_P}{\hbar}\frac{ Mm}{r^{d-1}}=  -\frac{G_d Mm}{r^{d-1}} 
\eeq
in $d$ spatial dimensions, 
\beq\label{Gd}
G_d = 2\pi^{1-\frac{d}{2}} \Gamma\left(\frac{d}{2}\right) \frac{c^3 \ell^{d-1}_P}{\hbar}
\eeq
is the $d$-dimensional gravitational constant. In this context we see that $\ell_P$ is the Planck length.

One can make further inferences regarding the nature of the screen.  The gravitational force law was derived with the only constraint on temperature being that it is a measure of the equipartition of energy on the screen.  An observer in the rest frame of the test mass $m$ will infer the existence of a temperature
\beq
T = \frac{\hbar a}{2\pi k_B c} 
\label{utemp}
\eeq
due to the Unruh effect \cite{unruh}, where $a$ is the acceleration of the test mass.  This can be taken to be the temperature of the screen, understood as the temperature required for $M$ to induce an acceleration $a$ on the test mass \cite{verlinde}.  From eq. (\ref{entrans}), eq. (\ref{entforce}) becomes 
\beq
F_{\rm entropic} = \frac{2\pi k_B T mc}{\hbar} = ma~~.
\eeq
or in other words the law of inertia. 
The entropy content of the screen can be inferred to be
\beq
S_{\rm screen} \sim N 
\eeq
which by the holographic assumption (\ref{Narea}) makes it proportional to the area of the screen 
in $d$ spatial dimensions.

\section{Applications to One-Dimensional Gravity}
\label{section2}

The preceding argument holds formally for all dimensions $d \geq 1$, despite the fact that in one spatial
dimension a screen is only a point and so has no area.  Furthermore, the Einstein tensor $G_{\mu\nu}$ is identically zero in two spacetime dimensions, making a connection with relativistic gravity somewhat problematic.   These observations motivate us to consider the formulation of entropic in $d=1$ as a separate case, to see what insights for emergent gravity might be gleaned.

The advent of lower-dimensional gravity yielded much insight into aspects of quantum gravity and relativistic physics (see \cite{collas,brown,henneaux,mannsik,mann1,mann2} for some expository introductions).    The richness of its content lies in the simplicity of the governing equations of motion.  The action cannot be that of the Einstein-Hilbert action, since the Ricci scalar is a topological invariant.  While it is common to adopt some general form of dilatonic gravity for the action, this generally yields a set of field equations whose metric dynamics are coupled with that of the dilaton.  It is possible, however, to obtain \cite{mann3} what might be regarded as the most straightforward exposition of the Einstein equation in two-dimensional spacetime 
\beq
R - \Lambda = 8\pi G_{1} T~~,
\label{eeq11}
\eeq
from the action
\beq
S[g_{\mu \nu}, \phi] = \int d^2x~\sqrt{-g} ( \psi R-\frac{1}{2}(\nabla \psi)^2 +{\cal L}_m-2\Lambda) 
\eeq
where $\psi$ is a scalar field and ${\cal L}_m$ is the matter Lagrangian.   Requiring a vanishing trace of the resulting stress-energy tensor decouples the dilaton from the background and recovers the desired vacuum spacetime structure. This theory has the unique feature that it reduces to Newtonian gravity in 2 spacetime dimensions \cite{2dRT}.  Such an action can also be generalized to the case of a $(1+1)$-dimensional non-commutative geometry \cite{jrmpn}.

For a vacuum ($T = 0$) shows $R$ completely determines the Riemann tensor.  The surprising implication is that even a spacetime devoid of matter may still possess curvature provided the cosmological constant is nonzero \cite{collas}.   In the presence of energy (matter), a number of black hole and event horizon solutions are possible \cite{mann1,mann2}, which can possess either attractive or repulsive properties.  For arbitrary cosmological constant $\Lambda$, the vacuum solution to Equation~\ref{eeq11} is \cite{mann2}
\beq
ds^2 = -\left(\mp \half |\Lambda| x^2 - 2G_{1}M |x| -C\right) dt^2 + \frac{dx^2}{\left(\mp \half | \Lambda |x^2 - 2G_{1}M|x| -C\right)}
\label{11metric}
\eeq
where the parameter $M$ corresponds to the ADM mass, and $C<0$ is a arbitrary constant whose value determines the causal structure of the spacetime \cite{manndan}.  The sign convention is deSitter ($-$) and anti-deSitter ($+$).  For $M>0$ the above metric describes the 2-dimensional analogue of a Schwarzschild black hole  \cite{mann2}.
A two-dimensional Riessner-Nordstr\"om spacetime with point charge $Q$ can be shown to mimic the above metric with the equivalence $|\Lambda| = Q^2/4$ \cite{mann4}. 

In the absence of a cosmological constant, two point masses $M$ and $m$ will experience an attractive gravitational force, where the mass separation $x$ as a function of proper time $\tau$ is governed by the familiar inertial equation \cite{mann2}
\beq
x(\tau) = -\frac{M}{2} \tau^2 \epsilon(x) + v_0\tau + x_0~~,~~\epsilon(x) = \theta(x) - \theta(-x)
\eeq
Note that this expression is manifestly position independent, as the function $\epsilon(x)$ reflects only the relative position of $m$ to $M$ ($v_0$ and $x_0$ are constants of integration).   Unlike classical Newtonian gravity in three spatial dimensions, the (1+1)-gravitational acceleration between masses is constant, and therefore so is the force law.  This is further evident from the form of (\ref{11metric}), whose $g_{00}$ component is linear in $|x|$, and thus by association so is the gravitational potential $\phi(x) = G_{1}M|x|$.
 
Another key novelty of two-dimensional black hole solutions is the existence of a gravitational temperature \cite{mann2}, whose value can be calculated via a Wick rotation (\ref{11metric}),
\beq
ds^2 = \alpha(x)~d\tau^2 + \alpha^{-1}(x)~dx^2 ~~\longrightarrow ~~ \alpha(x(r))~d\tau^2 + dr^2~~,\alpha(x) = \left(\frac{dx}{dr}\right)^2
\eeq
The periodicity of $\alpha$ yields the standard temperature at the horizon $x_H$
\beq
T = \frac{\hbar}{2\pi} \left|\frac{\alpha^\prime(x_H)}{2}\right| = \frac{\hbar}{2\pi}\sqrt{M^2 - C\frac{\Lambda}{2}}
\eeq
where the latter equality follows upon using (\ref{11metric}).
In contrast to (3+1)- and higher dimensional gravity, it has been shown \cite{mann2} that the temperature of these black holes scales with its mass. 

It follows from foundational principles that if these solutions exhibit a Hawking temparture, an associated Beckenstein-Hawking entropy must also be a feature of 1+1-dimensional models.  It is thus natural to extend (or subtend) the (3+1)-formalism of entropic gravity to lower dimensional spacetimes, since 
as noted in \cite{verlinde}, the formalism may be extended to any higher-dimensional space ({\it i.e.} $d=3+n$ dimensions) thanks to the generalizability of the holographic principle, $S \sim A_{n+3}/4$.  While the Newtonian potential varies spatially as $\phi(r) \sim r^{-(n+1)}$ for $n>0$, and thus the force as $F(r) \sim r^{-(d+2)}$, their characteristics are notably different when $d < 3$. 

As demonstrated above  the one-dimensional spatial potential is linear in $x$, thus it should be possible to derive an associated entropic force.  The derivation for such is identical to that of Verlinde,  except that it is halted short of invoking the relationship between bit density $N$ and area $A$ (and implicitly spatial separation $r$) due to dimensional limitations.  The general entropic force expression is thus
\beq\label{Fentgen}
F_{\rm entropic} = -T\frac{\Delta S}{\Delta x} = -mM \left(\frac{4\pi c^3}{N\hbar}\right)~~.
\eeq
upon using eqs. (\ref{equip}) and (\ref{entrans}), with $E= M c^2 = \frac{N}{2} k_b T$.  
Note that the expression (\ref{Fentgen}) does not require the use of (\ref{Narea}) and
is valid in {\it any} number of dimensions. 

The spatial dependence of the entropic force is introduced implicitly though the dependence of $N$ on $A(r)$ via eq. (\ref{Narea}), which at first suggests that the method cannot be implemented in linear space.  The above relation suggests, however, that the fundamental quantity of interest is the bit-count $N$, and not the bounding area.  As a result, in one spatial dimension we have
\beq
F_{1} = G_{1} mM~~,
\eeq
which is in agreement with the well-known result that the gravitational force in one dimension is constant, provided the coupling itself is defined as
\beq
G_{1} = \left(\frac{4\pi c^3}{N\hbar}\right)~~.
\label{g11}
\eeq
 implying that  
\beq\label{N1d}
N = 4\pi c^3/G_1\hbar
\eeq
expressing the number of bits in terms of the speed of light, Planck's constant, and $G$.  Using Eq.~ (\ref{Gd}) the $d=1$ coupling is $G_1 =2\pi c^3/\hbar$, one can obtain a numerical value for the number of bits $N$ that reside at each point in the spacetime.  Since Eq.~(\ref{Narea}) gives a value of $A=2$ in this case -- {\i.e.} both antipodal points $(-x,x)$ comprising the boundary of the 1-volume -- we divide the resulting calculation by two and conclude that 
 \beq\label{N1d}
N = 1~~.
\eeq
That is, there is one bit of information at every point in the two-dimensional spacetime.  We note that the relation between $G$ and the generalized Planck area in different dimensions has also been considered in References \cite{klink1} and \cite{klink2}.  The case of $(1+1)$-dimensions was not considered, however, so our result may be compared for consistency to the general cases discussed therein. 

The implications for entropic gravity are profound.   Since the right-hand-side of eq. (\ref{N1d}) is a constant, the immediate implication is that $N$ itself is also constant.  That is, unlike in higher dimensional spaces, the number of bits contained at a point in (1-D) space is constant, irrespective of the distance from the mass $M$ that generates the screen.   

This is a novel result that is specific to the $(1+1)$ framework, albeit one that contains several subtleties that merit discussion.  First, the original Verlinde argument relies on temperature, a quantity that traditionally is associated in the Maxwellian sense with vibrational modes of a system.  A similar argument could be extended to the temperature on a black hole horizon.   In our context, the temperature is that associated with the degrees of freedom located at a point, which henceforth must be considered as  internal degrees of freedom.   A related argument pertains to the applicability of energy equipartition at a point.

An additional point to highlight is the dependence of the coupling on bit density.  The result of equation (\ref{g11}) is applicable in arbitrary dimensions, implying   $N \sim G^{-1}$.  This yields the potentially counter-intuitive conclusion that more information implies weaker gravity, and vise versa.  In the generalized model, however, the bit density depends on the area of the screen, $N \sim A$.  Larger screen areas imply larger $N$, and consequently this leads to a system with weak gravitation.  The two points can be reconciled, nevertheless, by interpreting that $N$ implies the possible number of {\it total} states that may result.  From this perspective, systems with large $N$ are very ``disordered'', a characteristic one would anticipate from a theory with weak gravity.  Likewise, a large value of $G$ stems from a small $N$, which simply implies a very strongly-coupled system with few degrees of freedom.  

This conclusion is in general agreement with tthat of Reference \cite{Paulos:2011zu}, which through constructing a holographic $c$-function that applies to black holes solutions as well as renormalization group flow backgrounds, derives a relation between the number of degrees of freedom and Wald entropy $\mathcal N=S_{\rm Wald}/\Omega$.  In this case, $\Omega$ is a measure of the holographic phase space volume -- {\it i.e.} the number of states -- which in our formulation corresponds to $N$.  From this relationship, Verlinde's entropic gravity can be rigorously derived from Einstein's equations.  Our above interpretation of $N$ as a state count and the argument of \cite{Paulos:2011zu} are thus equivalent.

A recent pedagogical analysis of the entropic gravity formalism \cite{sabine} suggested that the framework is applicable in spaces of any dimension $d$ (which is of course true), except the ``pathological'' case of $d=1$.  The reason given is that area is not defined for $d<2$.  As we have demonstrated, however, this point can be rendered moot if one stops short of associating the number of bits with an area.  

Indeed, the thrust of our conclusions lies in the interplay between entropy, area, number of bits, and gravitation, as exemplified in Equation~(\ref{g11}).  This implicit relationship between $G, N,$ and $A$ is also highlighted in~\cite{sabine}, where it is noted therein that since $A \sim G N$, the number of bits may actually be excluded from the calculation since $N$ is determined by the values of $A$ and $G$.   We note that this relationship breaks down in one dimension, however, since $A_1 \rightarrow 0$, suggesting that entropic gravity is ill-defined for this manifold.  In the general formulation for entropic gravity, the appearance of $A$ emerges from $N\equiv N(r) = A(r)/\ell_P^{d-1}$.  Our definition (\ref{g11}) does not necessitate the introduction of an area, if one accepts the notion that $N$ is an internal parameter.
 
Furthermore, one may use this fact to  reinterpret the very notion of area.  Since $A$ does not enter the entropic gravity formulation directly except by substitution for the bit count $N(r) \rightarrow A(r)$, one may regard there to be no formal difference between $d=1$ and $d > 1$.  That is, each case depends fundamentally on $N$, and not on area.   In this sense, area is an emergent quality of how information permeates a space.  In $1-D$, it concentrates at a point, but in $d > 1$ spaces it evenly distributes itself in space and thus defines the ``area" as $A_d(r) = N  \ell_p^{d-1}$.

\section{Concluding Remarks and Future Considerations}\label{concspec}

We have provided a comprehensive treatment of $(1+1)-$dimensional entropic gravity in the classical limit.  The formalism correctly reproduces the expected constant gravitational law, which has the profound implication that the information density at any point along the line must be constant.  Furthermore, we have introduced a new interpretation of area as a uniform distribution of entropy in $d-$dimensions.  This allows for a natural extension of the entropy gravity mechanism to one spatial dimension, where ``area'' does not exist.  

Based on these conclusions, it is tempting to postulate a relativistic extension to this mechanism.  Area -- and hence entropy -- is a natural characteristic of spacetime foliations.  The union of such objects thus defines the complete manifold, from which Einstein's equations may be extracted in the same spirit as Newton's laws.  Geodesics are then the extremal flow lines of entropy.  Future work on these ideas, and thus a model of emergent entropic general relativity, is thus warranted. 

\vskip 2cm

\noindent{\bf Acknowledgments}\\
 We thank Prasanna Bhogale for interesting discussions.   RBM was financially supported by NSERC and JRM by the Research Corporation for Science Advancement.  JRM would additionally like to acknowledge the generous hospitality of the University of Waterloo Department of Physics and Astronomy and the Perimeter Institute for Theoretical Physics, at which this research was conducted.

 \end{document}